\documentclass[%
superscriptaddress,
preprint,
 amsmath,amssymb,
 aps,
]{revtex4-1}

\usepackage{graphicx}
\graphicspath{{img/}}
\usepackage{caption}
\usepackage{subcaption}
\usepackage{floatrow}

\usepackage{dcolumn}
\usepackage{bm}
\usepackage[hidelinks]{hyperref}
\usepackage[mathlines]{lineno}

\begin{document}


\title{Temperature Dependent \textit{n-p} Transition of 3 Dimensional Dirac Semimetal Na$_3$Bi Thin Film}

\author{Chang Liu}
 \affiliation{School of Physics and Astronomy, Monash University, Victoria 3800 Australia}
  \affiliation{Monash Centre for Atomically Thin Materials, Monash University, Victoria 3800 Australia}
 \affiliation{
 ARC Centre of Excellence in Future Low-Energy Electronics Technologies,\\
 Monash University, Victoria 3800 Australia
 }
 
\author{Jack Hellerstedt}%
 \affiliation{School of Physics and Astronomy, Monash University, Victoria 3800 Australia}%
 \affiliation{Institute of Physics of the Czech Academy of Sciences, v.v.i., Cukrovarnicka 10, 162 00 Prague, Czech Republic}

\author{Mark T. Edmonds}
 \affiliation{School of Physics and Astronomy, Monash University, Victoria 3800 Australia}%
 \affiliation{Monash Centre for Atomically Thin Materials, Monash University, Victoria 3800 Australia}
 \affiliation{
 ARC Centre of Excellence in Future Low-Energy Electronics Technologies,\\
 Monash University, Victoria 3800 Australia
 }

\author{Michael S. Fuhrer*}
 \affiliation{School of Physics and Astronomy, Monash University, Victoria 3800 Australia}%
 \affiliation{Monash Centre for Atomically Thin Materials, Monash University, Victoria 3800 Australia}
 \affiliation{
 ARC Centre of Excellence in Future Low-Energy Electronics Technologies,\\
 Monash University, Victoria 3800 Australia
 }

\date{\today}

\begin{abstract}
We study the temperature dependence ($77$ K - $475$ K) of the longitudinal resistivity and Hall coefficient of thin films (thickness $20$ nm) of three dimensional topological Dirac semimetal Na$_3$Bi grown via molecular beam epitaxy (MBE). The temperature-dependent Hall coefficient is electron-like at low temperature, but transitions to hole-like transport around $200$ K. We develop a model of a Dirac band with electron-hole asymmetry in Fermi velocity and mobility (assumed proportional to the square of Fermi velocity) which explains well the magnitude and temperature dependence of the Hall resistivity. We find that the hole mobility is about $7$ times larger than the electron mobility. In addition, we find that the electron mobility decreases significantly with increasing temperature, suggesting electron-phonon scattering strongly limits the room temperature mobility.\\
\newline
\newline
* Corresponding author: michael.fuhrer@monash.edu

\end{abstract}

\maketitle

Research into two-dimensional ($2$D) and three-dimensional ($3$D) Dirac materials in condensed matter physics has attracted much interest since the experimental isolation of graphene \cite{novoselov2005two,geim2007rise}. Na$_{3}$Bi is a topological Dirac semimetal (TDS) with stable 3D Dirac points separated form the $\Gamma$ point along the $k_z$ axis in its 1st Brillouin zone which has a linear band dispersion in all three momentum space directions \cite{wang2012dirac,liu2014discovery}. The behavior of Dirac fermions in $3$D opens up the possibility to study phenomena such as the chiral anomaly \cite{xiong2015evidence,burkov2016z}; thin films geometry offers the opportunity to study ambipolar electronic devices \cite{hellerstedt2017electrostatic}, electron-hole pudding \cite{edmonds2016spatial}, along with the theoretical prediction of opening a band gap in excess of $300$ meV in monolayer films \cite{niu2017robust} and a topological phase transition between trivial and non-trivial insulator that occurs at varying thickness or with applied electric field \cite{pan2015electric,xiao2015anisotropic}.

Recently, Na$_3$Bi thin films grown on sapphire and silicon dioxide substrates have been studied at low temperature ($5$ K) to reveal electron doped films with carrier mobility in excess of $6000$ cm$^2$V$^{-1}$s$^{-1}$ and a Hall carrier density lower than $1.0\times10^{18}$ cm$^{-3}$ \cite{hellerstedt2016electronic}. Modulation of the carrier density via molecular doping (sapphire) \cite{edmonds2016molecular} or backgate (SiO$_2$) \cite{hellerstedt2017electrostatic} has been used to reach the charge neutrality point.

Yet little is known about the temperature dependent transport properties of Na$_3$Bi thin films. Such information is crucial for low-doped films, to understand whether the close approach to the Dirac point is limited by spatial variations in the Fermi energy $E_F$ due to disorder \cite{adam2007self,skinner2014coulomb,ramakrishnan2015transport} or thermal excitation. In graphene on SiO$_2$, the magnitude of the fluctuations is on order $100$ meV, with characteristic Fermi temperatures $T_F = E_F/k_B > 1,000$ K, where $k_B$ is the Boltzmann constant \cite{adam2007self}. Hence most experiments are performed in the regime $T\ll T_F$, where thermal excitation of electron-hole pairs near the Dirac point is generally unimportant. Cleaner graphene has enabled a closer approach to the Dirac point and observation of new physics in the regime $T \geq T_F$ \cite{crossno2016observation,bandurin2016negative}. Due to better electronic screening in $3$D, spatial fluctuations in Fermi energy in $3$D TDS are expected to be much smaller for a comparable disorder. Indeed, low fluctuations in $E_F$ of $4\sim6$ meV $=$ $40\sim70$ K were observed in thin film Na$_3$Bi, suggesting the regime $T\geq T_F$ is experimentally accessible \cite{edmonds2016spatial}.

Here we measure the temperature-dependent longitudinal resistivity $\rho_{xx}(T)$  and Hall coefficient $R_H(T)$ of low-doped thin-film Na$_3$Bi. $R_H(T)$ is negative (\textit{n}-type) at low temperature, relatively temperature-independent, and reflects a carrier density of $1.0$-$1.3$ $\times$ $10^{19}$ cm$^{-3}$ which changes with thermal annealing of the thin film. On raising the temperature, $R_H$ transitions to positive (\textit{p}-type), and at the highest temperatures shows a magnitude and temperature dependence independent of annealing conditions. We explain the results using a simple model based on the two-carrier transport of electrons and holes, with an asymmetry in electron and hole Fermi velocity and hence mobility (assumed proportional to the square of Fermi velocity). The model indicates a ratio of electron to hole velocity (mobility) of $0.37$ ($0.14$). The higher velocity and mobility for holes is unusual for a conventional semiconductor but reasonable for the inverted bands of a TDS.

A major practical obstacle in the study of Na$_3$Bi thin films is their extreme reactivity to ambient.  We avoid this problem by using a home built mini molecular beam epitaxy (MBE) chamber, specifically equipped for measurement of temperature- and magnetic field- dependent transport properties of thin films grown via MBE in a Hall bar geometry defined \emph{in-situ} using a stencil mask \cite{hellerstedt2013situ}. Film growth was performed on an atomically flat insulating $\alpha$-Al$_2$O$_3$[$0001$] substrate (Shinkosha Japan) pre-patterned with Ti/Au electrodes ($5/50$ nm) prior to introduction to UHV. A two-step growth method was employed, similar to previous thin film growths on sapphire \cite{hellerstedt2016electronic}, with the first $2$ nm grown at $115$ $^{\circ}$C and remaining $18$ nm grown at a final temperature of $240$ $^{\circ}$C in a $10$:$1$ Na:Bi ratio. Post growth resistivity and Hall effect were measured between $77$ K and $475$ K.

\begin{figure*}[!h]
    \centering
    \captionsetup{justification=justified,width=0.9\textwidth}
    \begin{subfigure}{0.48\textwidth}
    \includegraphics[width=1\linewidth]{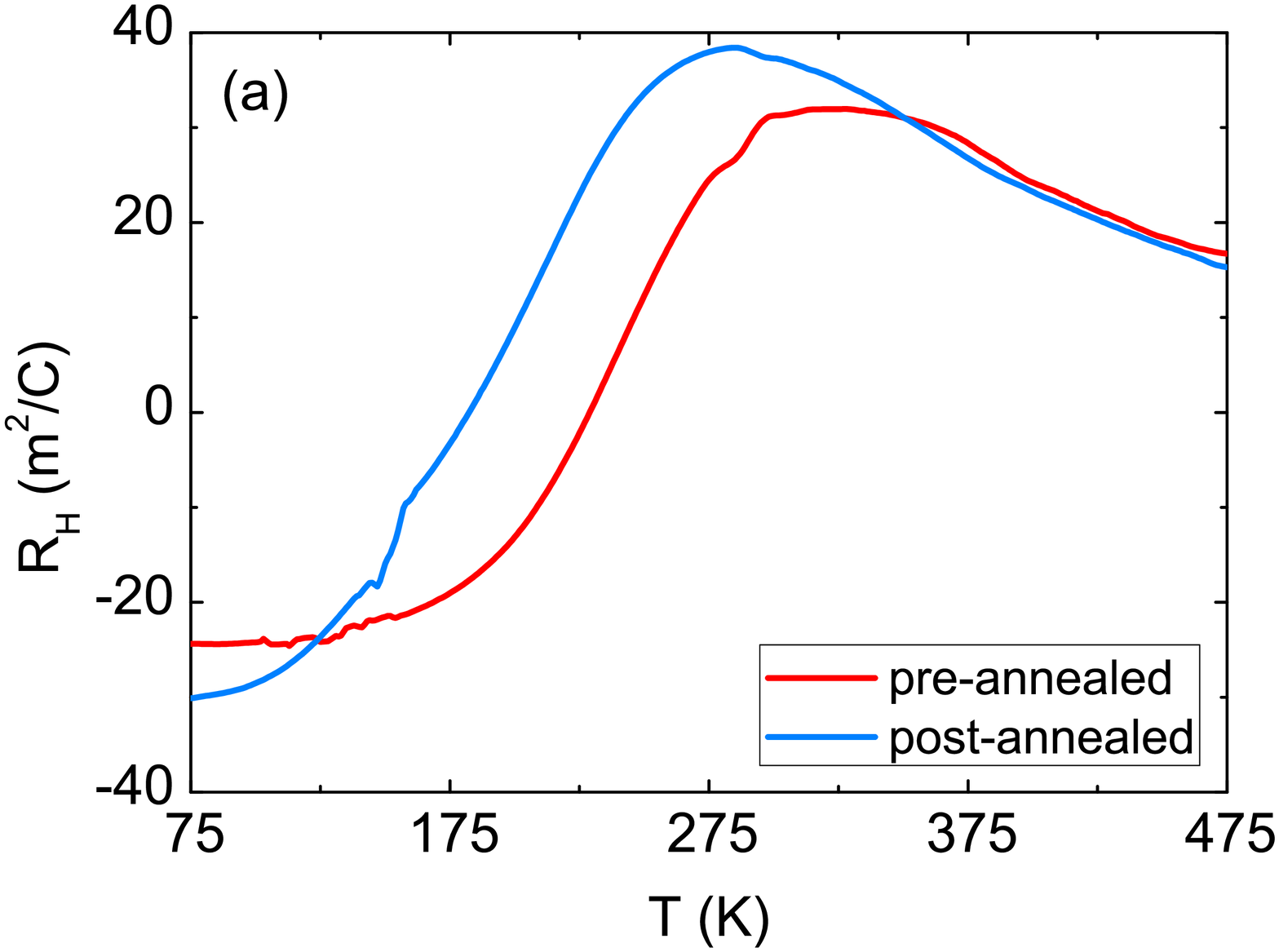} 
    \end{subfigure}
    \begin{subfigure}{0.48\textwidth}
    \includegraphics[width=1\linewidth]{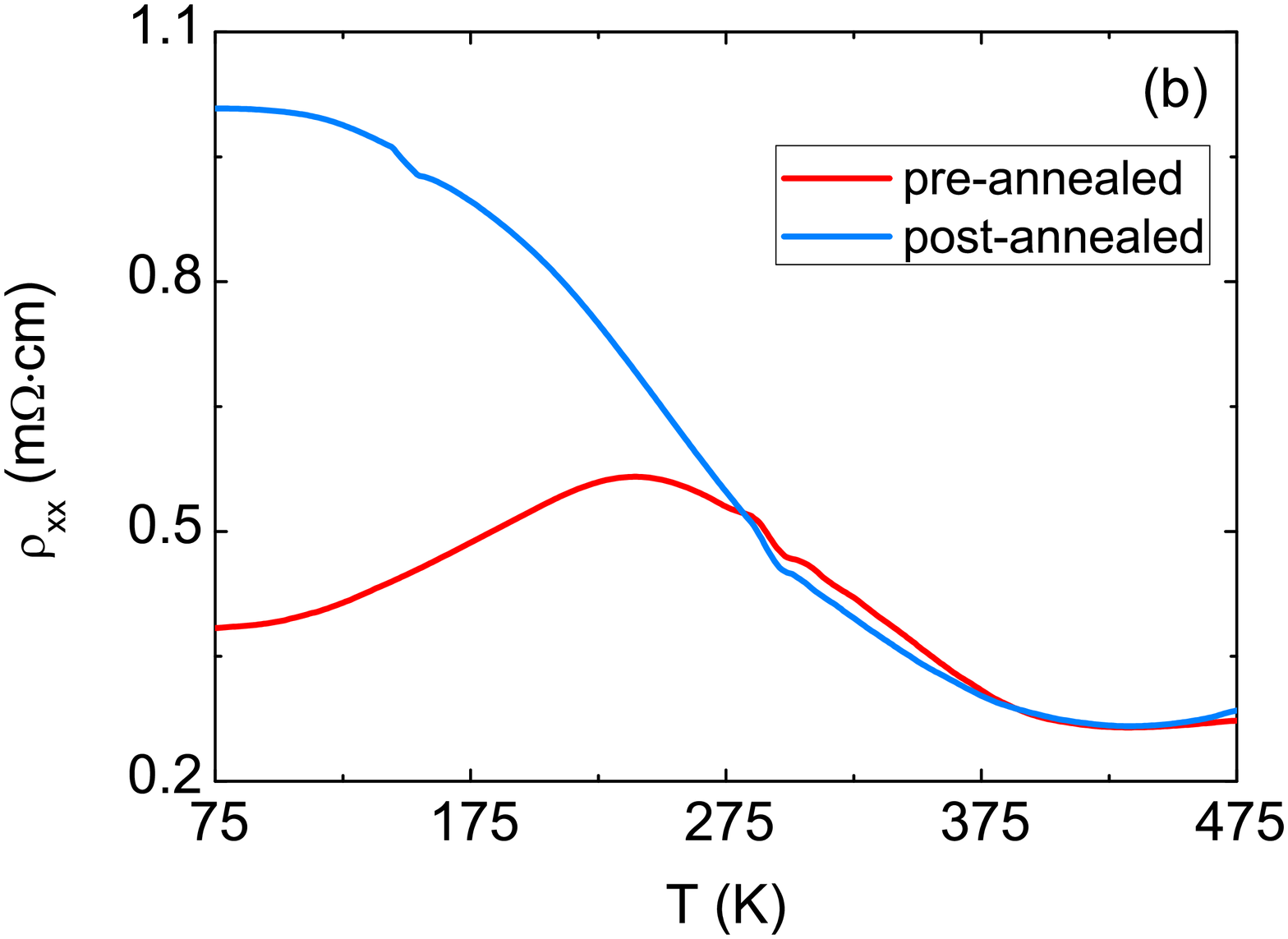} 
    \end{subfigure}
    \caption{(Color online) Temperature-dependent-transport measurements on a $20$ nm Na$_3$Bi thin film before (red) and after (blue) an annealing cycle. (a) Temperature-dependent Hall coefficient $R_H(T)$. (b) Temperature-dependent resistivity $\rho_{xx}(T)$.}
    \label{fig_Experiment_result}
\end{figure*}

Figure \ref{fig_Experiment_result} shows the measured Hall coefficient $R_H$ and  longitudinal resistivity $\rho_{xx}$ of a Na$_3$Bi thin film before (pre-annealed) and after (post-annealed) a one-hour annealing cycle at $473$ K. Figure \ref{fig_Experiment_result}(a) shows that at low temperature the sample is electron-like (i.e. $R_H$ is negative) with $R_H$ weakly dependent on temperature between $77$-$175$ K, but dependent on annealing conditions. The pre-annealed sample (red) shows $R_H$ saturated at about  $-24$ m$^2$C$^{-1}$ at $77$ K corresponding to a Hall carrier density of $-2.6\times10^{13}$ cm$^{-2}$, while the post-annealed sample (blue) shows $R_H = -30$ m$^2$C$^{-1}$ (Hall carrier density $-2.1\times10^{13}$ cm$^{-2}$). Upon raising the temperature $R_H$ becomes hole-like (i.e. $R_H$ is positive) below room temperature with a \textit{n-p} transition temperature that varies for the pre-annealed and post-annealed sample. At the highest temperatures, $R_H$ becomes independent of annealing condition and decreases with temperature. Figure \ref{fig_Experiment_result}(b) shows the temperature-dependent longitudinal resistivity. At high temperature, the pre-annealed and post-annealed films show very similar resistivity, and the negative slope is typical of semiconductors in which thermal activation of carriers dominates the temperature-dependent resistivity. At low temperature, the pre-annealed film (exhibiting higher low-temperature carrier density) shows lower resistivity with metallic temperature dependence, while the post-annealed film shows higher resistivity and semiconducting behavior.

A perfectly electron-hole symmetric Dirac band should not exhibit such a temperature-dependent \textit{n-p} transition. We therefore consider a simplified model of an asymmetric band by introducing an asymmetry of the Fermi velocity, $v_{e}$ in the conduction band and $v_{h}$ in the valence band. We assume $v_h = 2.43\times 10^{5}$ m$\cdot$s$^{-1}$ which is the geometric mean value of the Fermi velocities in all three directions ($k_x$, $k_y$ and $k_z$) from ARPES experiments \cite{liu2014discovery}. We further assume that the ratio of electron to hole mobility is proportional to the square of the ratio of Fermi velocities \cite{sarma2015carrier}, resulting in a single asymmetry parameter:

\begin{equation}
    b=\frac{\mu_{e}}{\mu_{h}}=\frac{v_{e}^2}{v_{h}^2}
    \label{eq_mobility_ratio}
\end{equation}
In the experiment, the sample is electron-like at low temperature and a low Fermi energy is inferred. As the temperature becomes comparable to the Fermi temperature, thermal activation of holes becomes important. The fact that $R_H$ switches sign is difficult to understand unless the holes have higher mobility, e.g. $b<1$. Hence we assumed an electron-hole mobility ratio $b\in(0,1)$.

\begin{figure*}[ht!]
    \centering
    \captionsetup{justification=justified, width=0.9\textwidth}
    \begin{subfigure}{0.45\textwidth}
    \includegraphics[width=1\linewidth]{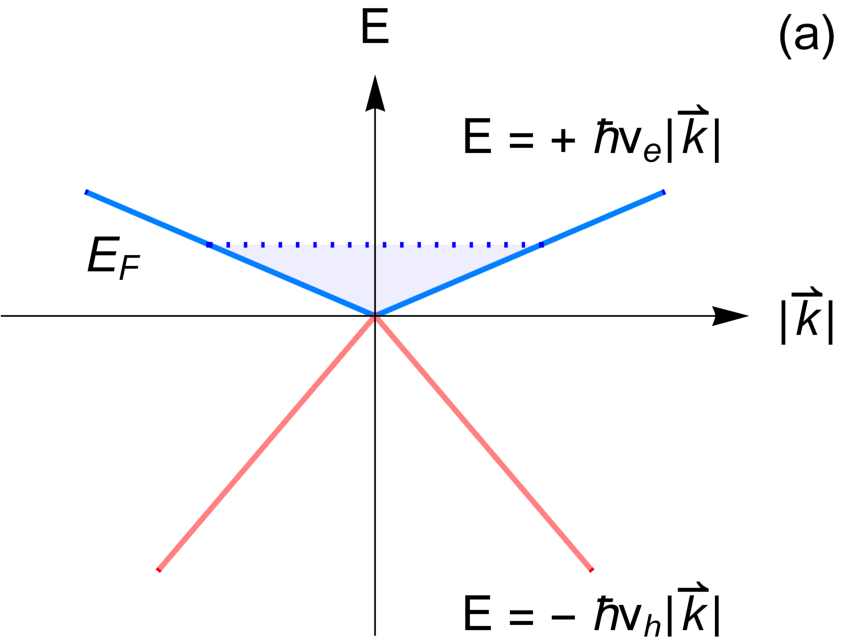} 
    \end{subfigure}
    \begin{subfigure}{0.45\textwidth}
    \includegraphics[width=1\linewidth]{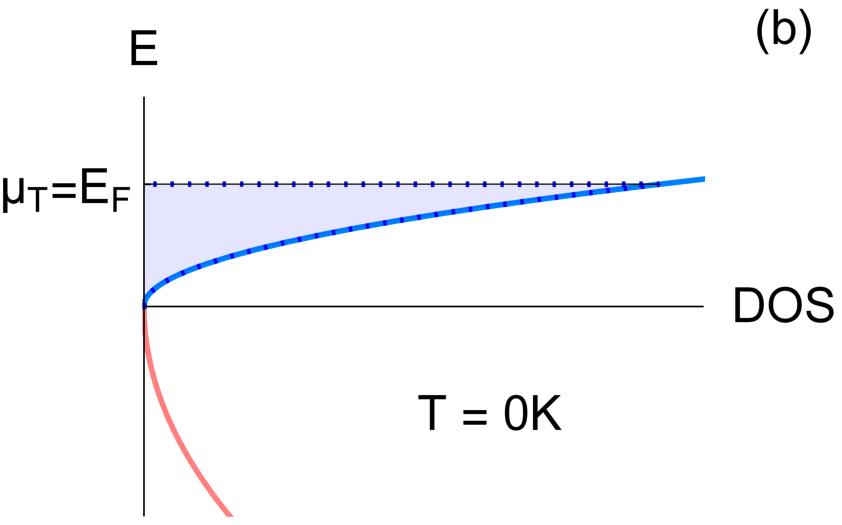} 
    \end{subfigure}
    \begin{subfigure}{0.45\textwidth}
    \includegraphics[width=1\linewidth]{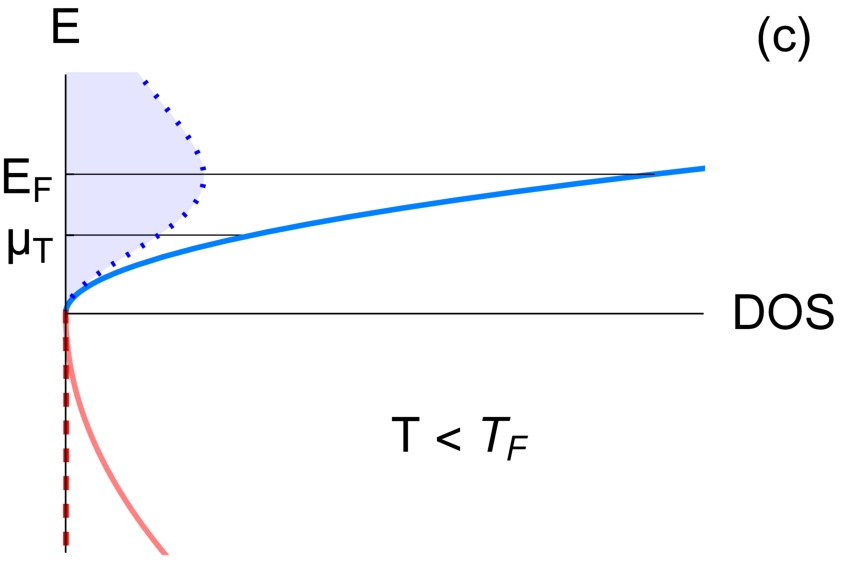} 
    \end{subfigure}
    \begin{subfigure}{0.45\textwidth}
    \includegraphics[width=1\linewidth]{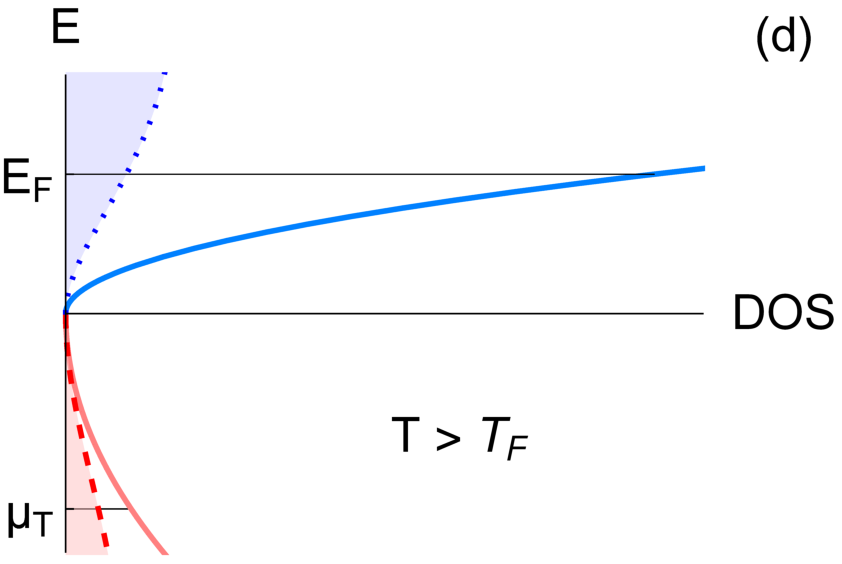} 
    \end{subfigure}
    \caption{Illustration of thermal activation in an asymmetric Dirac band structure used in the model. (a) The Dirac cone with different Fermi velocity in each band which is higher in valence band. (b-d) The respective density of states (DOS) and the occupation of the Dirac band structure, at zero temperature (b), low temperature compared to Fermi temperature $T < T_F$ (c), and high temperature $T > T_F$ (d).}
    \label{fig_model}
\end{figure*}

Figure \ref{fig_model} illustrates the situation schematically. Figure \ref{fig_model}(a) shows an asymmetric Dirac cone with extra electron doping. With the assumption of the energy dispersion $E=+v_{e}\hbar|\vec{k}|$ in the conduction band and $E=-v_{h}\hbar|\vec{k}|$ in valence band where $\hbar$ is the Planck constant, $E$ is the energy respect to Dirac point. The relative density of states (DOS) of such a Dirac cone is also asymmetric, $DOS(E) = \frac{g}{2\pi^2}(\frac{1}{\hbar v_{F}})^3 E^2$, where $g = 4$ is the spin/valley degeneracy and $E$ is the energy with respect to the Dirac point, as shown in Figure \ref{fig_model}(b). Figure 2(b) represents the zero temperature case, where the chemical potential $\mu$ (Fermi level) is located at the Fermi energy ($E_F$) above the Dirac point due to electron doping. In this case the hole band is fully filled, with carriers in the electron band the only charge carrier with a total number density of $N_{E_{F}} = \frac{g}{6\pi^2}(\frac{E_F}{\hbar v_e})^3$. As the temperature increases, thermal activation is governed by Fermi-Dirac statistic which gives the electron and hole concentrations

\begin{equation}
n(T) = -\frac{g}{\pi^2}\left(\frac{k_B T}{\hbar v_e}\right)^3 \text{Li}_{3}\left[-e^{\frac{\mu}{k_B T}}\right]
\label{eq_electron_density}
\end{equation}

\begin{equation}
p(T) = -\frac{g}{\pi^2}\left(\frac{k_B T}{\hbar v_h}\right)^3 \text{Li}_{3}\left[-e^{\frac{-\mu}{k_B T}}\right]
\label{eq_hole_density}
\end{equation}
where Li$_3(x)$ is the Polylogarithm of order three. These quantities are subject to the addition constraint of charge density conservation, $n(T) = N_{E_{F}} + p(T)$. When $T<T_F$ (see Figure \ref{fig_model}(c)), the electron distribution in the conduction band is broadened while there are few holes generated in the valence band and the transport is still dominated by electrons. At a higher temperature $T>T_F$ (Figure \ref{fig_model}(d)), holes contribute more to current with further thermal excitation. Therefore, it is clear that even though the sample remains net electron doped, holes may dominate the transport if the hole mobility is much larger than the electron mobility.

\begin{figure*}[ht!]
    \centering
    \captionsetup{justification=justified, width=0.9\textwidth}
    \includegraphics[width=1\linewidth]{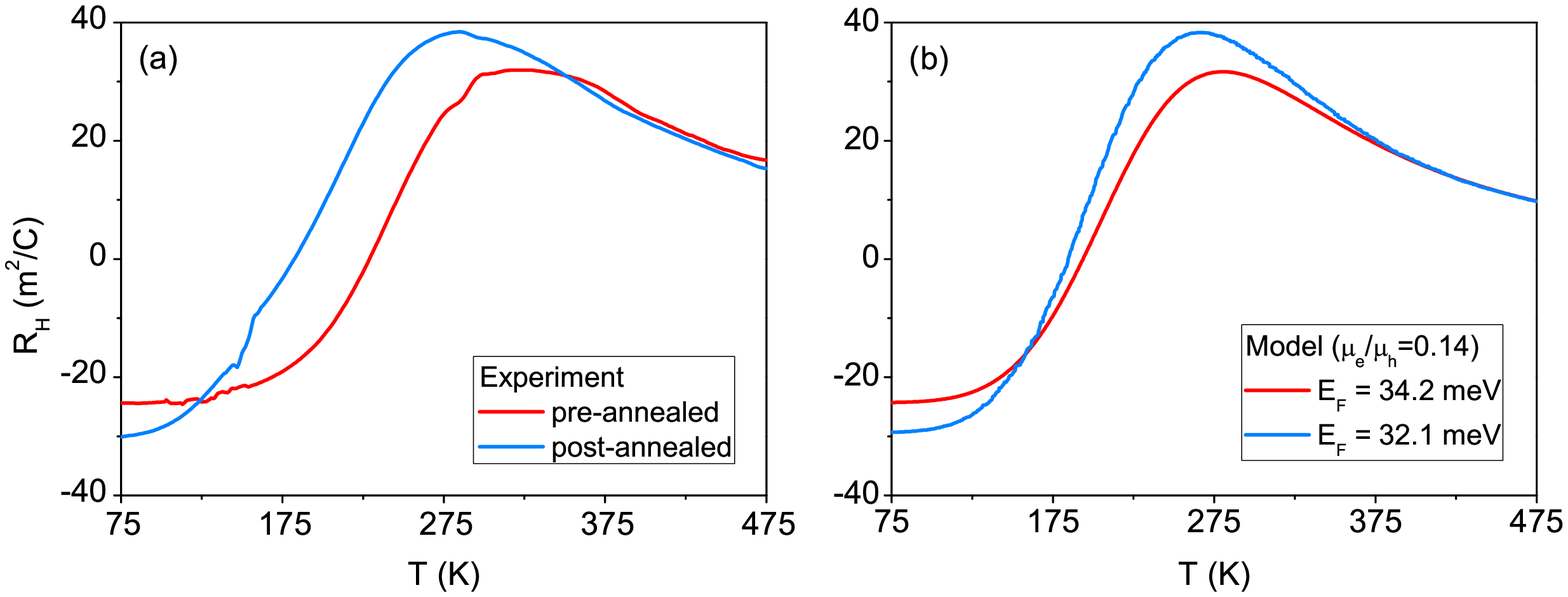} 
    
    \caption{(Color online) Comparison of the temperature-dependent Hall coefficient in (a) experiment and (b) simulation where an electron-hole mobility ratio $b = 0.14$ was used for two different Fermi energy case, $34.2$ meV (red) and $32.1$ meV (blue).}

    \label{fig_RH_vs_T_Exp_Sim}
\end{figure*}

Figure \ref{fig_RH_vs_T_Exp_Sim}(b) shows the results of the model. We fit the data by assuming a global asymmetry parameter $b$, and further, assume $E_F$ depends on the annealing condition. We find the results are well described by a mobility ratio $b = 0.14$, and Fermi energies $E_F = 34.2$ meV pre-annealed (red) and $32.1$ meV post-annealed (blue) as shown in Figure \ref{fig_RH_vs_T_Exp_Sim}(b). The model explains well the experimental observations, where at low temperatures the Hall coefficient largely reflects the $E_F$ at each annealing condition. This agrees with the expectation that an \textit{n}-doped Na$_3$Bi film has larger negative Hall coefficient at low temperature when the Fermi energy is lower. The temperature dependence is stronger, i.e. begins at lower temperatures, for the lower-doped annealed film (blue) compared to the pre-annealed film (red); the lower $E_F$ means thermal activation is important at a lower temperature. At the highest temperatures, the model reproduces the magnitude and temperature dependence seen in the experiment, including the $E_F$-independent magnitude of $R_H$.

\begin{figure*}[ht!]
    \centering
    \captionsetup{justification=justified, width=0.9\textwidth}
    \includegraphics[width=1\linewidth]{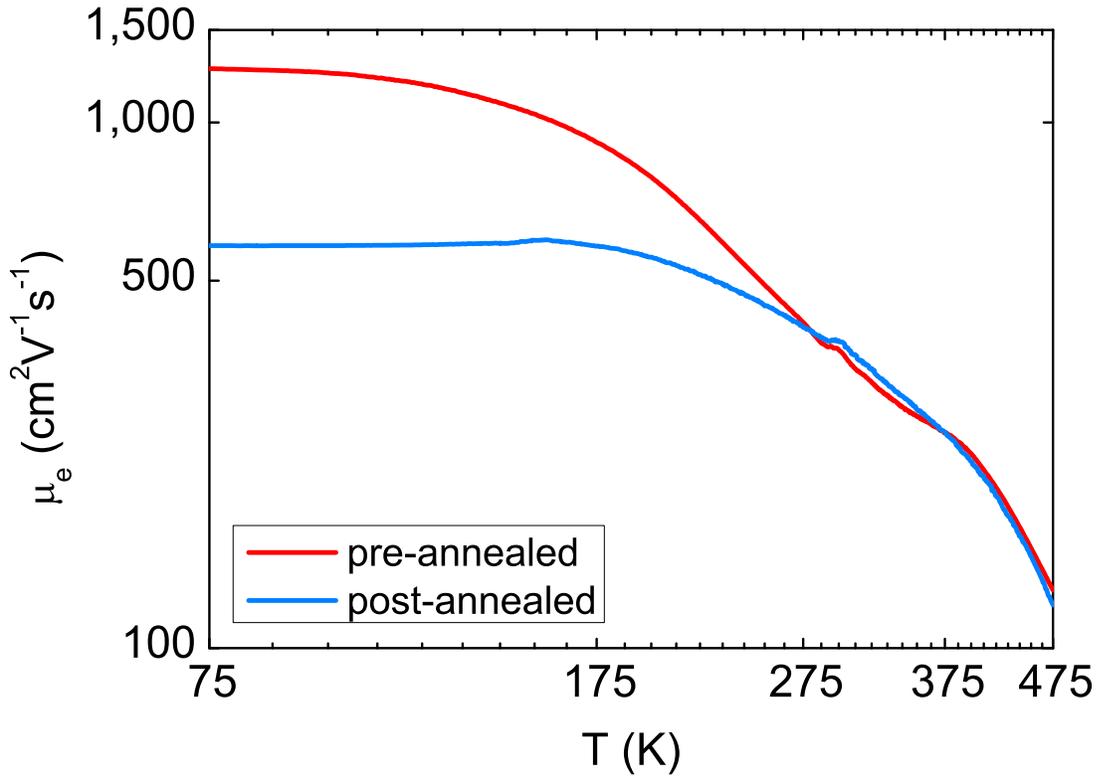} 
    \caption{(Color online) Electron mobility versus temperature for the Na$_3$Bi thin film before annealing (red) and after annealing (blue).}
    \label{fig_mobility_T}
\end{figure*}

We note that $R_H(T)$ depends only on the ratio of mobilities $b$, which we assume is constant, while $\rho_{xx}(T)$ carries information about the magnitude of the mobilities. Hence, because our model predicts the electron and hole concentrations, we can also extract the temperature-dependent mobility from the $\rho_{xx}(T)$. Figure \ref{fig_mobility_T} shows the electron mobility as a function of temperature on a log-log scale for the pre-annealed (red) and post-annealed (blue) film. It is clear that for both cases $\mu_e$ saturates in the low-temperature range $T < 175$ K and decrease between $175$ K $ < T < 475$ K. In the highest temperature range $T > 275$ K, the mobility is independent of annealing conditions. The temperature dependence is strong, with $\mu_e \sim T^a$ with $a\in[2,4]$.

We consider two possible causes of the temperature-dependent mobility: temperature-dependent screening of static disorder, and electron-phonon scattering. Temperature-dependent screening in a $3$D TDS is expected to give a conductivity $\sigma \sim T^4$ at high temperature  $T>T_F$ \cite{rodionov2015conductivity}, while the total carrier density goes as $n(T) + p(T) \sim T^3$. Hence the mobility is expected to \textit{increase} with increasing temperature, in contradiction to our findings. Thus we conclude that electron-phonon scattering dominates the temperature dependent mobility in our samples. This explains our observation of $R_H$, $\rho_{xx}$ and $\mu_e$ independent of Fermi energy and disorder at high temperature, which is expected in the regime $T > T_F$ where the carrier density is dominated by thermal activation, if the mobility is dominated by electron-phonon scattering (i.e. independent of disorder). Surprisingly the mobility in this regime is only $\sim 300$ cm$^2$/Vs at room temperature, much lower than other Dirac systems limited by electron-phonon scattering \cite{kim2012intrinsic,zhu2009carrier}. The low mobility magnitude and strong temperature dependence may imply that low-frequency optical phonon modes in Na$_3$Bi are important scatterers \cite{jenkins2016three}. The strong electron-phonon scattering also indicates that lower temperatures and Fermi energies are needed to observe effects of temperature-dependent screening.

In summary, we have demonstrated that Na$_3$Bi thin films display a temperature dependent \textit{n-p} transition, where the films are \textit{n}-type at low temperature ($<175$ K) and \textit{p}-type at room temperature. This \textit{n-p} transition was accurately modeled by using a simplified Dirac band structure with higher hole band Fermi velocity than that of the electron band and an assumption of hole mobility 7 times greater than the electron mobility. A similar mobility asymmetry was also recently inferred from gated-tuned Na$_3$Bi transport measurements \cite{hellerstedt2017electrostatic} in good agreement with our results. Thermal activation in an asymmetric Dirac band also likely explains the \textit{n-p} transition in low-doped bulk Na$_3$Bi crystals \cite{xiong2015evidence}. Our study allows us to extract the temperature dependence of electron mobility which decreases strongly with increasing temperature which we interpret as due to strong electron-phonon scattering, limiting the mobility to $\sim 300$ cm$^2$/Vs at room temperature.

CL, MTE and MSF acknowledge support from the ARC Centre of Excellence in Future Low-Energy Electronics Technologies (CE170100039). JH, CL and M.S.F. are supported by the ARC under FL120100038. MTE acknowledges funding from DE160101157. We also acknowledge funding from the Monash Centre for Atomically Thin Materials Research Support Scheme.  This work was performed in part at the Melbourne Centre for Nanofabrication (MCN) in the Victorian Node of the Australian National Fabrication Facility (ANFF).

\medskip
\newpage
\addcontentsline{toc}{chapter}{References}
\clearpage
\bibliography{reference}

\begin{thebibliography}{24}%
\makeatletter
\providecommand \@ifxundefined [1]{%
 \@ifx{#1\undefined}
}%
\providecommand \@ifnum [1]{%
 \ifnum #1\expandafter \@firstoftwo
 \else \expandafter \@secondoftwo
 \fi
}%
\providecommand \@ifx [1]{%
 \ifx #1\expandafter \@firstoftwo
 \else \expandafter \@secondoftwo
 \fi
}%
\providecommand \natexlab [1]{#1}%
\providecommand \enquote  [1]{``#1''}%
\providecommand \bibnamefont  [1]{#1}%
\providecommand \bibfnamefont [1]{#1}%
\providecommand \citenamefont [1]{#1}%
\providecommand \href@noop [0]{\@secondoftwo}%
\providecommand \href [0]{\begingroup \@sanitize@url \@href}%
\providecommand \@href[1]{\@@startlink{#1}\@@href}%
\providecommand \@@href[1]{\endgroup#1\@@endlink}%
\providecommand \@sanitize@url [0]{\catcode `\\12\catcode `\$12\catcode
  `\&12\catcode `\#12\catcode `\^12\catcode `\_12\catcode `\%12\relax}%
\providecommand \@@startlink[1]{}%
\providecommand \@@endlink[0]{}%
\providecommand \url  [0]{\begingroup\@sanitize@url \@url }%
\providecommand \@url [1]{\endgroup\@href {#1}{\urlprefix }}%
\providecommand \urlprefix  [0]{URL }%
\providecommand \Eprint [0]{\href }%
\providecommand \doibase [0]{http://dx.doi.org/}%
\providecommand \selectlanguage [0]{\@gobble}%
\providecommand \bibinfo  [0]{\@secondoftwo}%
\providecommand \bibfield  [0]{\@secondoftwo}%
\providecommand \translation [1]{[#1]}%
\providecommand \BibitemOpen [0]{}%
\providecommand \bibitemStop [0]{}%
\providecommand \bibitemNoStop [0]{.\EOS\space}%
\providecommand \EOS [0]{\spacefactor3000\relax}%
\providecommand \BibitemShut  [1]{\csname bibitem#1\endcsname}%
\let\auto@bib@innerbib\@empty
\bibitem [{\citenamefont {Novoselov}\ \emph {et~al.}(2005)\citenamefont
  {Novoselov}, \citenamefont {Geim}, \citenamefont {Morozov}, \citenamefont
  {Jiang}, \citenamefont {Katsnelson}, \citenamefont {Grigorieva},
  \citenamefont {Dubonos},\ and\ \citenamefont {Firsov}}]{novoselov2005two}%
  \BibitemOpen
  \bibfield  {author} {\bibinfo {author} {\bibfnamefont {K.~S.}\ \bibnamefont
  {Novoselov}}, \bibinfo {author} {\bibfnamefont {A.~K.}\ \bibnamefont {Geim}},
  \bibinfo {author} {\bibfnamefont {S.}~\bibnamefont {Morozov}}, \bibinfo
  {author} {\bibfnamefont {D.}~\bibnamefont {Jiang}}, \bibinfo {author}
  {\bibfnamefont {M.}~\bibnamefont {Katsnelson}}, \bibinfo {author}
  {\bibfnamefont {I.}~\bibnamefont {Grigorieva}}, \bibinfo {author}
  {\bibfnamefont {S.}~\bibnamefont {Dubonos}}, \ and\ \bibinfo {author}
  {\bibfnamefont {A.}~\bibnamefont {Firsov}},\ }\href@noop {} {\bibfield
  {journal} {\bibinfo  {journal} {Nature}\ }\textbf {\bibinfo {volume} {438}},\
  \bibinfo {pages} {197} (\bibinfo {year} {2005})}\BibitemShut {NoStop}%
\bibitem [{\citenamefont {Geim}\ and\ \citenamefont
  {Novoselov}(2007)}]{geim2007rise}%
  \BibitemOpen
  \bibfield  {author} {\bibinfo {author} {\bibfnamefont {A.~K.}\ \bibnamefont
  {Geim}}\ and\ \bibinfo {author} {\bibfnamefont {K.~S.}\ \bibnamefont
  {Novoselov}},\ }\href@noop {} {\bibfield  {journal} {\bibinfo  {journal}
  {Nat. Mater.}\ }\textbf {\bibinfo {volume} {6}},\ \bibinfo {pages} {183}
  (\bibinfo {year} {2007})}\BibitemShut {NoStop}%
\bibitem [{\citenamefont {Wang}\ \emph {et~al.}(2012)\citenamefont {Wang},
  \citenamefont {Sun}, \citenamefont {Chen}, \citenamefont {Franchini},
  \citenamefont {Xu}, \citenamefont {Weng}, \citenamefont {Dai},\ and\
  \citenamefont {Fang}}]{wang2012dirac}%
  \BibitemOpen
  \bibfield  {author} {\bibinfo {author} {\bibfnamefont {Z.}~\bibnamefont
  {Wang}}, \bibinfo {author} {\bibfnamefont {Y.}~\bibnamefont {Sun}}, \bibinfo
  {author} {\bibfnamefont {X.-Q.}\ \bibnamefont {Chen}}, \bibinfo {author}
  {\bibfnamefont {C.}~\bibnamefont {Franchini}}, \bibinfo {author}
  {\bibfnamefont {G.}~\bibnamefont {Xu}}, \bibinfo {author} {\bibfnamefont
  {H.}~\bibnamefont {Weng}}, \bibinfo {author} {\bibfnamefont {X.}~\bibnamefont
  {Dai}}, \ and\ \bibinfo {author} {\bibfnamefont {Z.}~\bibnamefont {Fang}},\
  }\href@noop {} {\bibfield  {journal} {\bibinfo  {journal} {Phys. Rev. B}\
  }\textbf {\bibinfo {volume} {85}},\ \bibinfo {pages} {195320} (\bibinfo
  {year} {2012})}\BibitemShut {NoStop}%
\bibitem [{\citenamefont {Liu}\ \emph {et~al.}(2014)\citenamefont {Liu},
  \citenamefont {Zhou}, \citenamefont {Zhang}, \citenamefont {Wang},
  \citenamefont {Weng}, \citenamefont {Prabhakaran}, \citenamefont {Mo},
  \citenamefont {Shen}, \citenamefont {Fang}, \citenamefont {Dai} \emph
  {et~al.}}]{liu2014discovery}%
  \BibitemOpen
  \bibfield  {author} {\bibinfo {author} {\bibfnamefont {Z.}~\bibnamefont
  {Liu}}, \bibinfo {author} {\bibfnamefont {B.}~\bibnamefont {Zhou}}, \bibinfo
  {author} {\bibfnamefont {Y.}~\bibnamefont {Zhang}}, \bibinfo {author}
  {\bibfnamefont {Z.}~\bibnamefont {Wang}}, \bibinfo {author} {\bibfnamefont
  {H.}~\bibnamefont {Weng}}, \bibinfo {author} {\bibfnamefont {D.}~\bibnamefont
  {Prabhakaran}}, \bibinfo {author} {\bibfnamefont {S.-K.}\ \bibnamefont {Mo}},
  \bibinfo {author} {\bibfnamefont {Z.}~\bibnamefont {Shen}}, \bibinfo {author}
  {\bibfnamefont {Z.}~\bibnamefont {Fang}}, \bibinfo {author} {\bibfnamefont
  {X.}~\bibnamefont {Dai}},  \emph {et~al.},\ }\href@noop {} {\bibfield
  {journal} {\bibinfo  {journal} {Science}\ }\textbf {\bibinfo {volume}
  {343}},\ \bibinfo {pages} {864} (\bibinfo {year} {2014})}\BibitemShut
  {NoStop}%
\bibitem [{\citenamefont {Xiong}\ \emph {et~al.}(2015)\citenamefont {Xiong},
  \citenamefont {Kushwaha}, \citenamefont {Liang}, \citenamefont {Krizan},
  \citenamefont {Hirschberger}, \citenamefont {Wang}, \citenamefont {Cava},\
  and\ \citenamefont {Ong}}]{xiong2015evidence}%
  \BibitemOpen
  \bibfield  {author} {\bibinfo {author} {\bibfnamefont {J.}~\bibnamefont
  {Xiong}}, \bibinfo {author} {\bibfnamefont {S.~K.}\ \bibnamefont {Kushwaha}},
  \bibinfo {author} {\bibfnamefont {T.}~\bibnamefont {Liang}}, \bibinfo
  {author} {\bibfnamefont {J.~W.}\ \bibnamefont {Krizan}}, \bibinfo {author}
  {\bibfnamefont {M.}~\bibnamefont {Hirschberger}}, \bibinfo {author}
  {\bibfnamefont {W.}~\bibnamefont {Wang}}, \bibinfo {author} {\bibfnamefont
  {R.}~\bibnamefont {Cava}}, \ and\ \bibinfo {author} {\bibfnamefont
  {N.}~\bibnamefont {Ong}},\ }\href@noop {} {\bibfield  {journal} {\bibinfo
  {journal} {Science}\ }\textbf {\bibinfo {volume} {350}},\ \bibinfo {pages}
  {413} (\bibinfo {year} {2015})}\BibitemShut {NoStop}%
\bibitem [{\citenamefont {Burkov}\ and\ \citenamefont
  {Kim}(2016)}]{burkov2016z}%
  \BibitemOpen
  \bibfield  {author} {\bibinfo {author} {\bibfnamefont {A.~A.}\ \bibnamefont
  {Burkov}}\ and\ \bibinfo {author} {\bibfnamefont {Y.~B.}\ \bibnamefont
  {Kim}},\ }\href@noop {} {\bibfield  {journal} {\bibinfo  {journal} {Phys.
  Rev. Lett.}\ }\textbf {\bibinfo {volume} {117}},\ \bibinfo {pages} {136602}
  (\bibinfo {year} {2016})}\BibitemShut {NoStop}%
\bibitem [{\citenamefont {Hellerstedt}\ \emph {et~al.}()\citenamefont
  {Hellerstedt}, \citenamefont {Yudhistira}, \citenamefont {Edmonds},
  \citenamefont {Liu}, \citenamefont {Collins}, \citenamefont {Adam},\ and\
  \citenamefont {Fuhrer}}]{hellerstedt2017electrostatic}%
  \BibitemOpen
  \bibfield  {author} {\bibinfo {author} {\bibfnamefont {J.}~\bibnamefont
  {Hellerstedt}}, \bibinfo {author} {\bibfnamefont {I.}~\bibnamefont
  {Yudhistira}}, \bibinfo {author} {\bibfnamefont {M.~T.}\ \bibnamefont
  {Edmonds}}, \bibinfo {author} {\bibfnamefont {C.}~\bibnamefont {Liu}},
  \bibinfo {author} {\bibfnamefont {J.}~\bibnamefont {Collins}}, \bibinfo
  {author} {\bibfnamefont {S.}~\bibnamefont {Adam}}, \ and\ \bibinfo {author}
  {\bibfnamefont {M.~S.}\ \bibnamefont {Fuhrer}},\ }\href@noop {} {\bibinfo
  {journal} {arXiv:1707.09286}\ }\BibitemShut {NoStop}%
\bibitem [{\citenamefont {Edmonds}\ \emph {et~al.}()\citenamefont {Edmonds},
  \citenamefont {Collins}, \citenamefont {Hellerstedt}, \citenamefont
  {Yudhistira}, \citenamefont {Gomes}, \citenamefont {Rodrigues}, \citenamefont
  {Adam},\ and\ \citenamefont {Fuhrer}}]{edmonds2016spatial}%
  \BibitemOpen
\bibfield  {journal} {  }\bibfield  {author} {\bibinfo {author} {\bibfnamefont
  {M.~T.}\ \bibnamefont {Edmonds}}, \bibinfo {author} {\bibfnamefont {J.~L.}\
  \bibnamefont {Collins}}, \bibinfo {author} {\bibfnamefont {J.}~\bibnamefont
  {Hellerstedt}}, \bibinfo {author} {\bibfnamefont {I.}~\bibnamefont
  {Yudhistira}}, \bibinfo {author} {\bibfnamefont {L.~C.}\ \bibnamefont
  {Gomes}}, \bibinfo {author} {\bibfnamefont {J.~N.}\ \bibnamefont
  {Rodrigues}}, \bibinfo {author} {\bibfnamefont {S.}~\bibnamefont {Adam}}, \
  and\ \bibinfo {author} {\bibfnamefont {M.~S.}\ \bibnamefont {Fuhrer}},\
  }\href@noop {} {\bibinfo  {journal} {arXiv:1612.03385}\ }\BibitemShut
  {NoStop}%
\bibitem [{\citenamefont {Niu}\ \emph {et~al.}(2017)\citenamefont {Niu},
  \citenamefont {Buhl}, \citenamefont {Bihlmayer}, \citenamefont {Wortmann},
  \citenamefont {Dai}, \citenamefont {Bl{\"u}gel},\ and\ \citenamefont
  {Mokrousov}}]{niu2017robust}%
  \BibitemOpen
\bibfield  {journal} {  }\bibfield  {author} {\bibinfo {author} {\bibfnamefont
  {C.}~\bibnamefont {Niu}}, \bibinfo {author} {\bibfnamefont {P.~M.}\
  \bibnamefont {Buhl}}, \bibinfo {author} {\bibfnamefont {G.}~\bibnamefont
  {Bihlmayer}}, \bibinfo {author} {\bibfnamefont {D.}~\bibnamefont {Wortmann}},
  \bibinfo {author} {\bibfnamefont {Y.}~\bibnamefont {Dai}}, \bibinfo {author}
  {\bibfnamefont {S.}~\bibnamefont {Bl{\"u}gel}}, \ and\ \bibinfo {author}
  {\bibfnamefont {Y.}~\bibnamefont {Mokrousov}},\ }\href@noop {} {\bibfield
  {journal} {\bibinfo  {journal} {Phys. Rev. B}\ }\textbf {\bibinfo {volume}
  {95}},\ \bibinfo {pages} {075404} (\bibinfo {year} {2017})}\BibitemShut
  {NoStop}%
\bibitem [{\citenamefont {Pan}\ \emph {et~al.}(2015)\citenamefont {Pan},
  \citenamefont {Wu}, \citenamefont {Liu},\ and\ \citenamefont
  {Yang}}]{pan2015electric}%
  \BibitemOpen
  \bibfield  {author} {\bibinfo {author} {\bibfnamefont {H.}~\bibnamefont
  {Pan}}, \bibinfo {author} {\bibfnamefont {M.}~\bibnamefont {Wu}}, \bibinfo
  {author} {\bibfnamefont {Y.}~\bibnamefont {Liu}}, \ and\ \bibinfo {author}
  {\bibfnamefont {S.~A.}\ \bibnamefont {Yang}},\ }\href@noop {} {\bibfield
  {journal} {\bibinfo  {journal} {Sci. Rep.}\ }\textbf {\bibinfo {volume} {5}}
  (\bibinfo {year} {2015})}\BibitemShut {NoStop}%
\bibitem [{\citenamefont {Xiao}\ \emph {et~al.}(2015)\citenamefont {Xiao},
  \citenamefont {Yang}, \citenamefont {Liu}, \citenamefont {Li},\ and\
  \citenamefont {Zhou}}]{xiao2015anisotropic}%
  \BibitemOpen
  \bibfield  {author} {\bibinfo {author} {\bibfnamefont {X.}~\bibnamefont
  {Xiao}}, \bibinfo {author} {\bibfnamefont {S.~A.}\ \bibnamefont {Yang}},
  \bibinfo {author} {\bibfnamefont {Z.}~\bibnamefont {Liu}}, \bibinfo {author}
  {\bibfnamefont {H.}~\bibnamefont {Li}}, \ and\ \bibinfo {author}
  {\bibfnamefont {G.}~\bibnamefont {Zhou}},\ }\href@noop {} {\bibfield
  {journal} {\bibinfo  {journal} {Sci. Rep.}\ }\textbf {\bibinfo {volume} {5}}
  (\bibinfo {year} {2015})}\BibitemShut {NoStop}%
\bibitem [{\citenamefont {Hellerstedt}\ \emph {et~al.}(2016)\citenamefont
  {Hellerstedt}, \citenamefont {Edmonds}, \citenamefont {Ramakrishnan},
  \citenamefont {Liu}, \citenamefont {Weber}, \citenamefont {Tadich},
  \citenamefont {O’Donnell}, \citenamefont {Adam},\ and\ \citenamefont
  {Fuhrer}}]{hellerstedt2016electronic}%
  \BibitemOpen
  \bibfield  {author} {\bibinfo {author} {\bibfnamefont {J.}~\bibnamefont
  {Hellerstedt}}, \bibinfo {author} {\bibfnamefont {M.~T.}\ \bibnamefont
  {Edmonds}}, \bibinfo {author} {\bibfnamefont {N.}~\bibnamefont
  {Ramakrishnan}}, \bibinfo {author} {\bibfnamefont {C.}~\bibnamefont {Liu}},
  \bibinfo {author} {\bibfnamefont {B.}~\bibnamefont {Weber}}, \bibinfo
  {author} {\bibfnamefont {A.}~\bibnamefont {Tadich}}, \bibinfo {author}
  {\bibfnamefont {K.~M.}\ \bibnamefont {O’Donnell}}, \bibinfo {author}
  {\bibfnamefont {S.}~\bibnamefont {Adam}}, \ and\ \bibinfo {author}
  {\bibfnamefont {M.~S.}\ \bibnamefont {Fuhrer}},\ }\href@noop {} {\bibfield
  {journal} {\bibinfo  {journal} {Nano Lett.}\ }\textbf {\bibinfo {volume}
  {16}},\ \bibinfo {pages} {3210} (\bibinfo {year} {2016})}\BibitemShut
  {NoStop}%
\bibitem [{\citenamefont {Edmonds}\ \emph {et~al.}(2016)\citenamefont
  {Edmonds}, \citenamefont {Hellerstedt}, \citenamefont {O’Donnell},
  \citenamefont {Tadich},\ and\ \citenamefont {Fuhrer}}]{edmonds2016molecular}%
  \BibitemOpen
  \bibfield  {author} {\bibinfo {author} {\bibfnamefont {M.~T.}\ \bibnamefont
  {Edmonds}}, \bibinfo {author} {\bibfnamefont {J.}~\bibnamefont
  {Hellerstedt}}, \bibinfo {author} {\bibfnamefont {K.~M.}\ \bibnamefont
  {O’Donnell}}, \bibinfo {author} {\bibfnamefont {A.}~\bibnamefont {Tadich}},
  \ and\ \bibinfo {author} {\bibfnamefont {M.~S.}\ \bibnamefont {Fuhrer}},\
  }\href@noop {} {\bibfield  {journal} {\bibinfo  {journal} {ACS Appl. Mater.
  Interfaces}\ }\textbf {\bibinfo {volume} {8}},\ \bibinfo {pages} {16412}
  (\bibinfo {year} {2016})}\BibitemShut {NoStop}%
\bibitem [{\citenamefont {Adam}\ \emph {et~al.}(2007)\citenamefont {Adam},
  \citenamefont {Hwang}, \citenamefont {Galitski},\ and\ \citenamefont
  {Sarma}}]{adam2007self}%
  \BibitemOpen
  \bibfield  {author} {\bibinfo {author} {\bibfnamefont {S.}~\bibnamefont
  {Adam}}, \bibinfo {author} {\bibfnamefont {E.}~\bibnamefont {Hwang}},
  \bibinfo {author} {\bibfnamefont {V.}~\bibnamefont {Galitski}}, \ and\
  \bibinfo {author} {\bibfnamefont {S.~D.}\ \bibnamefont {Sarma}},\ }\href@noop
  {} {\bibfield  {journal} {\bibinfo  {journal} {Proc. Natl. Acad. Sci.
  U.S.A.}\ }\textbf {\bibinfo {volume} {104}},\ \bibinfo {pages} {18392}
  (\bibinfo {year} {2007})}\BibitemShut {NoStop}%
\bibitem [{\citenamefont {Skinner}(2014)}]{skinner2014coulomb}%
  \BibitemOpen
  \bibfield  {author} {\bibinfo {author} {\bibfnamefont {B.}~\bibnamefont
  {Skinner}},\ }\href@noop {} {\bibfield  {journal} {\bibinfo  {journal} {Phys.
  Rev. B}\ }\textbf {\bibinfo {volume} {90}},\ \bibinfo {pages} {060202}
  (\bibinfo {year} {2014})}\BibitemShut {NoStop}%
\bibitem [{\citenamefont {Ramakrishnan}\ \emph {et~al.}(2015)\citenamefont
  {Ramakrishnan}, \citenamefont {Milletari},\ and\ \citenamefont
  {Adam}}]{ramakrishnan2015transport}%
  \BibitemOpen
  \bibfield  {author} {\bibinfo {author} {\bibfnamefont {N.}~\bibnamefont
  {Ramakrishnan}}, \bibinfo {author} {\bibfnamefont {M.}~\bibnamefont
  {Milletari}}, \ and\ \bibinfo {author} {\bibfnamefont {S.}~\bibnamefont
  {Adam}},\ }\href@noop {} {\bibfield  {journal} {\bibinfo  {journal} {Phys.
  Rev. B}\ }\textbf {\bibinfo {volume} {92}},\ \bibinfo {pages} {245120}
  (\bibinfo {year} {2015})}\BibitemShut {NoStop}%
\bibitem [{\citenamefont {Crossno}\ \emph {et~al.}(2016)\citenamefont
  {Crossno}, \citenamefont {Shi}, \citenamefont {Wang}, \citenamefont {Liu},
  \citenamefont {Harzheim}, \citenamefont {Lucas}, \citenamefont {Sachdev},
  \citenamefont {Kim}, \citenamefont {Taniguchi}, \citenamefont {Watanabe}
  \emph {et~al.}}]{crossno2016observation}%
  \BibitemOpen
  \bibfield  {author} {\bibinfo {author} {\bibfnamefont {J.}~\bibnamefont
  {Crossno}}, \bibinfo {author} {\bibfnamefont {J.~K.}\ \bibnamefont {Shi}},
  \bibinfo {author} {\bibfnamefont {K.}~\bibnamefont {Wang}}, \bibinfo {author}
  {\bibfnamefont {X.}~\bibnamefont {Liu}}, \bibinfo {author} {\bibfnamefont
  {A.}~\bibnamefont {Harzheim}}, \bibinfo {author} {\bibfnamefont
  {A.}~\bibnamefont {Lucas}}, \bibinfo {author} {\bibfnamefont
  {S.}~\bibnamefont {Sachdev}}, \bibinfo {author} {\bibfnamefont
  {P.}~\bibnamefont {Kim}}, \bibinfo {author} {\bibfnamefont {T.}~\bibnamefont
  {Taniguchi}}, \bibinfo {author} {\bibfnamefont {K.}~\bibnamefont {Watanabe}},
   \emph {et~al.},\ }\href@noop {} {\bibfield  {journal} {\bibinfo  {journal}
  {Science}\ }\textbf {\bibinfo {volume} {351}},\ \bibinfo {pages} {1058}
  (\bibinfo {year} {2016})}\BibitemShut {NoStop}%
\bibitem [{\citenamefont {Bandurin}\ \emph {et~al.}(2016)\citenamefont
  {Bandurin}, \citenamefont {Torre}, \citenamefont {Kumar}, \citenamefont
  {Shalom}, \citenamefont {Tomadin}, \citenamefont {Principi}, \citenamefont
  {Auton}, \citenamefont {Khestanova}, \citenamefont {Novoselov}, \citenamefont
  {Grigorieva} \emph {et~al.}}]{bandurin2016negative}%
  \BibitemOpen
  \bibfield  {author} {\bibinfo {author} {\bibfnamefont {D.}~\bibnamefont
  {Bandurin}}, \bibinfo {author} {\bibfnamefont {I.}~\bibnamefont {Torre}},
  \bibinfo {author} {\bibfnamefont {R.~K.}\ \bibnamefont {Kumar}}, \bibinfo
  {author} {\bibfnamefont {M.~B.}\ \bibnamefont {Shalom}}, \bibinfo {author}
  {\bibfnamefont {A.}~\bibnamefont {Tomadin}}, \bibinfo {author} {\bibfnamefont
  {A.}~\bibnamefont {Principi}}, \bibinfo {author} {\bibfnamefont
  {G.}~\bibnamefont {Auton}}, \bibinfo {author} {\bibfnamefont
  {E.}~\bibnamefont {Khestanova}}, \bibinfo {author} {\bibfnamefont
  {K.}~\bibnamefont {Novoselov}}, \bibinfo {author} {\bibfnamefont
  {I.}~\bibnamefont {Grigorieva}},  \emph {et~al.},\ }\href@noop {} {\bibfield
  {journal} {\bibinfo  {journal} {Science}\ }\textbf {\bibinfo {volume}
  {351}},\ \bibinfo {pages} {1055} (\bibinfo {year} {2016})}\BibitemShut
  {NoStop}%
\bibitem [{\citenamefont {Hellerstedt}\ \emph {et~al.}(2013)\citenamefont
  {Hellerstedt}, \citenamefont {Chen}, \citenamefont {Kim}, \citenamefont
  {Cullen}, \citenamefont {Zheng},\ and\ \citenamefont
  {Fuhrer}}]{hellerstedt2013situ}%
  \BibitemOpen
  \bibfield  {author} {\bibinfo {author} {\bibfnamefont {J.}~\bibnamefont
  {Hellerstedt}}, \bibinfo {author} {\bibfnamefont {J.}~\bibnamefont {Chen}},
  \bibinfo {author} {\bibfnamefont {D.}~\bibnamefont {Kim}}, \bibinfo {author}
  {\bibfnamefont {W.~G.}\ \bibnamefont {Cullen}}, \bibinfo {author}
  {\bibfnamefont {C.}~\bibnamefont {Zheng}}, \ and\ \bibinfo {author}
  {\bibfnamefont {M.~S.}\ \bibnamefont {Fuhrer}},\ }\href@noop {} {\bibfield
  {journal} {\bibinfo  {journal} {Proc. SPIE}\ }\textbf {\bibinfo {volume}
  {8923}},\ \bibinfo {pages} {89230P} (\bibinfo {year} {2013})}\BibitemShut
  {NoStop}%
\bibitem [{\citenamefont {Sarma}\ \emph {et~al.}(2015)\citenamefont {Sarma},
  \citenamefont {Hwang},\ and\ \citenamefont {Min}}]{sarma2015carrier}%
  \BibitemOpen
  \bibfield  {author} {\bibinfo {author} {\bibfnamefont {S.~D.}\ \bibnamefont
  {Sarma}}, \bibinfo {author} {\bibfnamefont {E.}~\bibnamefont {Hwang}}, \ and\
  \bibinfo {author} {\bibfnamefont {H.}~\bibnamefont {Min}},\ }\href@noop {}
  {\bibfield  {journal} {\bibinfo  {journal} {Phys. Rev. B}\ }\textbf {\bibinfo
  {volume} {91}},\ \bibinfo {pages} {035201} (\bibinfo {year}
  {2015})}\BibitemShut {NoStop}%
\bibitem [{\citenamefont {Rodionov}\ and\ \citenamefont
  {Syzranov}(2015)}]{rodionov2015conductivity}%
  \BibitemOpen
  \bibfield  {author} {\bibinfo {author} {\bibfnamefont {Y.~I.}\ \bibnamefont
  {Rodionov}}\ and\ \bibinfo {author} {\bibfnamefont {S.}~\bibnamefont
  {Syzranov}},\ }\href@noop {} {\bibfield  {journal} {\bibinfo  {journal}
  {Phys. Rev. B}\ }\textbf {\bibinfo {volume} {91}},\ \bibinfo {pages} {195107}
  (\bibinfo {year} {2015})}\BibitemShut {NoStop}%
\bibitem [{\citenamefont {Kim}\ \emph {et~al.}(2012)\citenamefont {Kim},
  \citenamefont {Li}, \citenamefont {Syers}, \citenamefont {Butch},
  \citenamefont {Paglione}, \citenamefont {Sarma},\ and\ \citenamefont
  {Fuhrer}}]{kim2012intrinsic}%
  \BibitemOpen
  \bibfield  {author} {\bibinfo {author} {\bibfnamefont {D.}~\bibnamefont
  {Kim}}, \bibinfo {author} {\bibfnamefont {Q.}~\bibnamefont {Li}}, \bibinfo
  {author} {\bibfnamefont {P.}~\bibnamefont {Syers}}, \bibinfo {author}
  {\bibfnamefont {N.~P.}\ \bibnamefont {Butch}}, \bibinfo {author}
  {\bibfnamefont {J.}~\bibnamefont {Paglione}}, \bibinfo {author}
  {\bibfnamefont {S.~D.}\ \bibnamefont {Sarma}}, \ and\ \bibinfo {author}
  {\bibfnamefont {M.~S.}\ \bibnamefont {Fuhrer}},\ }\href@noop {} {\bibfield
  {journal} {\bibinfo  {journal} {Phys. Rev. Lett.}\ }\textbf {\bibinfo
  {volume} {109}},\ \bibinfo {pages} {166801} (\bibinfo {year}
  {2012})}\BibitemShut {NoStop}%
\bibitem [{\citenamefont {Zhu}\ \emph {et~al.}(2009)\citenamefont {Zhu},
  \citenamefont {Perebeinos}, \citenamefont {Freitag},\ and\ \citenamefont
  {Avouris}}]{zhu2009carrier}%
  \BibitemOpen
  \bibfield  {author} {\bibinfo {author} {\bibfnamefont {W.}~\bibnamefont
  {Zhu}}, \bibinfo {author} {\bibfnamefont {V.}~\bibnamefont {Perebeinos}},
  \bibinfo {author} {\bibfnamefont {M.}~\bibnamefont {Freitag}}, \ and\
  \bibinfo {author} {\bibfnamefont {P.}~\bibnamefont {Avouris}},\ }\href@noop
  {} {\bibfield  {journal} {\bibinfo  {journal} {Phys. Rev. B}\ }\textbf
  {\bibinfo {volume} {80}},\ \bibinfo {pages} {235402} (\bibinfo {year}
  {2009})}\BibitemShut {NoStop}%
\bibitem [{\citenamefont {Jenkins}\ \emph {et~al.}(2016)\citenamefont
  {Jenkins}, \citenamefont {Lane}, \citenamefont {Barbiellini}, \citenamefont
  {Sushkov}, \citenamefont {Carey}, \citenamefont {Liu}, \citenamefont
  {Krizan}, \citenamefont {Kushwaha}, \citenamefont {Gibson}, \citenamefont
  {Chang} \emph {et~al.}}]{jenkins2016three}%
  \BibitemOpen
  \bibfield  {author} {\bibinfo {author} {\bibfnamefont {G.}~\bibnamefont
  {Jenkins}}, \bibinfo {author} {\bibfnamefont {C.}~\bibnamefont {Lane}},
  \bibinfo {author} {\bibfnamefont {B.}~\bibnamefont {Barbiellini}}, \bibinfo
  {author} {\bibfnamefont {A.}~\bibnamefont {Sushkov}}, \bibinfo {author}
  {\bibfnamefont {R.}~\bibnamefont {Carey}}, \bibinfo {author} {\bibfnamefont
  {F.}~\bibnamefont {Liu}}, \bibinfo {author} {\bibfnamefont {J.}~\bibnamefont
  {Krizan}}, \bibinfo {author} {\bibfnamefont {S.}~\bibnamefont {Kushwaha}},
  \bibinfo {author} {\bibfnamefont {Q.}~\bibnamefont {Gibson}}, \bibinfo
  {author} {\bibfnamefont {T.-R.}\ \bibnamefont {Chang}},  \emph {et~al.},\
  }\href@noop {} {\bibfield  {journal} {\bibinfo  {journal} {Phys. Rev. B}\
  }\textbf {\bibinfo {volume} {94}},\ \bibinfo {pages} {085121} (\bibinfo
  {year} {2016})}\BibitemShut {NoStop}%
\end{thebibliography}%

\newpage

\newpage

\newpage

\newpage

\end{document}